\documentclass[a4paper,twocolumn,nofootinbib]{revtex4}
\pdfoutput=1
\usepackage[colorlinks,linkcolor=darkblue,citecolor=darkblue,urlcolor=darkblue]{
hyperref}
\usepackage{graphicx}
\usepackage{tabularx}
\usepackage{array}
\usepackage{amsmath}
\usepackage{amssymb}
\usepackage{xcolor}
\usepackage{longtable}
\usepackage{verbatim}
\usepackage{bbm}
\usepackage{multirow}
\usepackage{colortbl}
\usepackage{slashed}
\usepackage[normalem]{ulem}
\definecolor{darkblue}{rgb}{0,0,0.5}
\allowdisplaybreaks

\begin{document}

\title{Violation of lepton flavour universality in composite Higgs models}
\author{Christoph Niehoff}
\email[E-mail: ]{christoph.niehoff@tum.de}
\affiliation{Excellence Cluster Universe, Technische Universit\"at M\"unchen, 
Boltzmannstr.~2, 85748~Garching, Germany}
\author{Peter Stangl}
\email[E-mail: ]{peter.stangl@tum.de}
\affiliation{Excellence Cluster Universe, Technische Universit\"at M\"unchen, 
Boltzmannstr.~2, 85748~Garching, Germany}
\author{David M. Straub}
\email[E-mail: ]{david.straub@tum.de}
\affiliation{Excellence Cluster Universe, Technische Universit\"at M\"unchen, 
Boltzmannstr.~2, 85748~Garching, Germany}

\begin{abstract}
\noindent
We investigate whether
the $2.6\sigma$ deviation from lepton flavour universality
in $B^+\to K^+\ell^+\ell^-$ decays recently observed at the LHCb experiment
can be explained in minimal composite Higgs models.
We show that a visible departure from universality is indeed possible
if left-handed muons have a sizable degree of compositeness.
Constraints from $Z$-pole observables are avoided by a custodial protection
of the muon coupling.
The deficit in the invisible $Z$ width at LEP is explained in the same
region of parameters.
\end{abstract}

\maketitle

\section{Introduction}

Rare $B$ meson decays based on the quark-level transition $b\to s\,\ell^+\ell^-$, 
with $\ell=e,\mu,\tau$, are sensitive probes of physics beyond the Standard 
Model (SM) as these flavour-changing neutral currents are loop and CKM 
suppressed in the SM. In addition to probing flavour-violation in the quark 
sector, also lepton flavour universality (LFU) can be tested by comparing the 
rates of processes with different leptons in the final state.
Recently, the LHCb Collaboration has measured the ratio $R_K$ of 
the $B^+\to K^+\mu^+\mu^-$ and $B^+\to K^+e^+e^-$ branching
ratios~\cite{Aaij:2014ora},
\begin{equation}
 R_K = \frac{\text{BR}(B^+ \to K^+\mu^+\mu^-)_{[1,6]}}{\text{BR}(B^+ \to 
K^+e^+e^-)_{[1,6]}} = 0.745^{+0.090}_{-0.074} \pm 0.036
 \,,
 \label{eq:RK}
\end{equation}
which corresponds to a $2.6\sigma$ deviation from the SM value,
which is $1.0$ to an excellent precision.
If confirmed, this deviation from unity would constitute an irrefutable evidence 
of new physics (NP).

Supposing the measurement \eqref{eq:RK} is indeed a sign of NP, it is 
interesting to ask which NP model could account for this sizable violation of 
LFU. It has been demonstrated already that in models where the $b\to 
s\,\ell^+\ell^-$ transition is mediated at the tree level by a heavy neutral gauge 
boson \cite{Altmannshofer:2014cfa,Buras:2014fpa,Glashow:2014iga,Bhattacharya:2014wla,
Crivellin:2015mga,Altmannshofer:2014rta,Crivellin:2015lwa} or by spin-0 or 
spin-1 leptoquarks 
\cite{Hiller:2014yaa,Biswas:2014gga,Buras:2014fpa,Sahoo:2015wya,Hiller:2014ula}, 
it is possible to explain the measurement without violating other constraints. 
However, in more complete models, it often turns out to be difficult to generate 
a large enough amount of LFU violation. In the MSSM, it has been shown that it 
is not possible to accommodate the central value of \eqref{eq:RK} 
\cite{Altmannshofer:2014rta}. In composite Higgs models, which at present 
arguably constitute the most compelling solution to the hierarchy problem next 
to supersymmetry, one interesting possibility recently considered to explain 
\eqref{eq:RK} is to postulate the presence of composite leptoquarks 
\cite{Gripaios:2014tna}. This however comes at the price of a significant 
complication of the models. In more minimal models a thorough analysis of the 
possible size of LFU violation is still lacking and it is the purpose of this 
study to fill this gap.

\section{FCNCs and partially composite muons}

A departure from LFU in $b\to s\,\ell^+\ell^-$ transitions can be described in the 
weak effective Hamiltonian by a non-universal shift in the Wilson coefficients 
of the operators
\begin{align}
O_9^{(\prime)\ell} &= 
(\bar{s} \gamma_{\mu} P_{L(R)} b)(\bar{\ell} \gamma^\mu \ell)\,
,\label{eq:O9}
\\
O_{10}^{(\prime)\ell} &=
(\bar{s} \gamma_{\mu} P_{L(R)} b)( \bar{\ell} \gamma^\mu \gamma_5 \ell)\,
.\label{eq:O10}
\end{align}
A global analysis has shown that the data prefer a negative shift in $C_9^\mu$, 
with a possible positive contribution to $C_{10}^\mu$ 
\cite{Altmannshofer:2014rta} (see also \cite{Ghosh:2014awa,Hurth:2014vma} for 
other recent fits). In the following, we will denote the shift in the Wilson 
coefficients with respect to their SM values by $\delta C_i$. Interestingly, for 
$\delta C_{10}^\mu=-\delta C_9^\mu$, which corresponds to the limit in which 
only the left-handed leptons are involved in the transition, a comparably good 
fit to the case of NP in $\delta C_9^\mu$ only is obtained.

In models with partial compositeness, there are two distinct tree-level 
contributions to the $b\to s\,\ell^+\ell^-$ transition 
(cf.~\cite{Straub:2013zca,Altmannshofer:2013foa}).
\begin{itemize}
\item $Z$ exchange, facilitated by a tree-level flavour-changing $Z$ coupling 
that arises from the mixing after EWSB of states with different $SU(2)_L$ 
quantum numbers; this effect is thus always parametrically suppressed by 
$v^2/f^2$, but not mass-suppressed.
\item Heavy neutral spin-1 resonance exchange. This effect does not require the 
insertion of a Higgs VEV, but is mass-suppressed by the heavy resonance 
propagator.
\end{itemize}
Concerning the heavy resonance exchange, one can distinguish two qualitatively 
different effects depending on how the coupling of the resonance to the 
final-state leptons comes about.
\begin{itemize}
 \item There is a contribution stemming from the mixing of the heavy resonances 
with the $Z$ boson; in this case, the coupling to the leptons is to a good 
approximation equal to the SM $Z$ coupling of the leptons.
 \item Another contribution stems from the mixing of the leptons with heavy 
vector-like composite leptons. While the coupling of the resonance to composite 
leptons is expected to be strong, this contribution is suppressed by the 
(squared) degree of compositeness of the light leptons.
\end{itemize}

A crucial observation first made in \cite{Altmannshofer:2013foa} is that both 
the $Z$-mediated contribution and the resonance exchange contribution based on 
vector boson mixing lead to $\delta C_9^\mu/\delta 
C_{10}^\mu=(1-4s_w^2)\approx0.08$ due to the (accidentally) small vector 
coupling of the $Z$ to charged leptons in the SM. Such a pattern of effects is 
not supported by the global fit to $b\to s$ data.

We are therefore led to the conclusion that the vector resonance exchange with 
the resonance-lepton coupling induced by the mixing of muons with heavy 
vector-like partners is the only way to explain the $R_K$ anomaly in our 
framework in accordance with the data.
While the product of degrees of compositeness of the left- and right-handed
muon needs to be small to account for the smallness of the muon mass,
one of the two could be sizable. In the case of left-handed muons, as mentioned
above, this would lead to a pattern $\delta C_9^\mu=-\delta C_{10}^\mu$,
while right-handed muons would imply $\delta C_9^\mu=+\delta C_{10}^\mu$.
The latter however is not preferred by the global fit to $b\to s$ data,
so we require the {\em left-handed muons} to be significantly composite.
The main questions are then:
\begin{itemize}
\item How large does the degree of compositeness of left-handed muons
have to be to explain \eqref{eq:RK}?
\item How large do precision measurements allow this degree of compositeness 
to be?
\end{itemize}

Concerning the first question, an important point is that the quark 
flavour-changing coupling to the vector resonances cannot be too large since it 
would otherwise lead to a large NP effect in $B_s$-$\bar B_s$ mixing that is not 
allowed by the data \cite{Altmannshofer:2014rta} (see also 
\cite{Altmannshofer:2009ma,Descotes-Genon:2013wba,Altmannshofer:2013foa,
Gauld:2013qba,Buras:2013qja}). Combining the $B_s$ mixing constraint with the 
requirement to get a visible effect in $R_K$ leads to a lower bound on the 
coupling of the vector resonances to muons. Estimating this coupling in our case 
as $g_\rho s_{L\mu}^2$, where $g_\rho$ is a generic (strong) coupling between 
the composite lepton partners and the vector resonances and $s_{L\mu}$ is the 
degree of compositeness of left-handed muons, and writing a common vector 
resonance mass as $m_\rho\equiv g_\rho f/2$,\footnote{\label{fn:f}Here,
$m_\rho\equiv g_\rho f/2$ is just a convenient definition because $f$ is the
suppression scale of dimension-6 operators mediated by vector resonance
exchange. In models with a composite pseudo-Goldstone boson Higgs, $f$
can be identified with the Goldstone boson's ``decay constant''.}
one finds that a visible effect in 
$R_K$ requires, up to a model-dependent $\mathcal{O}(1)$ factor, ${s_{L\mu}\gtrsim0.15\,\xi^{-1/4}}$
, where $\xi=v^2/f^2$.

Such a sizable degree of compositeness is problematic at first sight. In 
general, the left-handed muons mix after EWSB with composite states that have 
different $SU(2)_L$ quantum numbers. This leads to a shift in the $Z$ coupling 
to left-handed muons that is generically of the size $\delta 
g_{Z\mu\mu}^L\sim\xi s_{L\mu}^2$. Given the LEP precision measurements which 
require $|\delta g_{Z\mu\mu}^L|\lesssim 10^{-3}$ implies, again up to a 
model-dependent $\mathcal{O}(1)$ factor, $s_{L\mu}\lesssim0.03\,\xi^{-1/2}$. Even if just 
a rough estimate, this shows clearly that a model satisfying this naive 
estimates is not viable. However, it is well-known that models exist where 
certain couplings of the $Z$ boson do not receive any corrections at tree level 
due to discrete symmetries: in the same way as this {\em custodial protection} 
prevents the $Z\bar b_L b_L$ coupling from large corrections
\cite{Agashe:2006at},
the $Z\bar \mu_L\mu_L$ coupling could be protected \cite{Agashe:2009tu},
opening the possibility of 
significantly composite left-handed muons.

\section{Model setup}

Composite Higgs models generally allow for many possibilities in model building. 
To make our results less model-dependent, our guideline will be to use the simplest model including partial compositeness. 
Indeed as we will see, very much is already fixed by demanding compatibility with electroweak precision tests.

In general, composite Higgs models feature a SM-like elementary sector and a strongly interacting BSM sector with a global symmetry $H$. 
It is well-known that in order to avoid critical tree-level corrections to the $T$ parameter one has to impose custodial symmetry, which is most easily done by choosing $H = SO(4) \sim SU(2)_L \times SU(2)_R$. 
We further assume that the global symmetry in the composite sector contains a $U(1)_X$ such that hypercharge is given by $Y=T_{3R}+X$ where $T_{3R}$ is the third component of right-handed isospin. 

Under the paradigm of partial compositeness the elementary leptons $\chi$ mix linearly with fermionic composite operators $\mathcal{O}_\mathrm{comp}^{(\chi)}$ such that $ \mathcal{L}_\mathrm{mix} = \sum_\chi \bar{\chi} \mathcal{O}_\mathrm{comp}^{(\chi)} + \mbox{h.c.}\,$.
Demanding a custodial protection of the $Z \bar{\mu}_L\mu_L$ vertex by the introduction of a discrete $P_{LR}$ symmetry restricts the possible choices for representations of the composite operators under the custodial symmetry \cite{Agashe:2006at}.
We find that for the operator mixing with the left-handed lepton doublet, this leaves only one possibility,
$(\mathbf{2}, \mathbf{2})_0$ under $SU(2)_L \times SU(2)_R \times U(1)_X$. 
By the same reasoning the right-handed muon then has to mix with a $(\mathbf{1}, \mathbf{3})_0$.
On the composite side we thus embed the lepton partners into a representation $(\mathbf{2}, \mathbf{2})_0 \oplus (\mathbf{1}, \mathbf{3})_0 \oplus (\mathbf{3}, \mathbf{1})_0$, where the $(\mathbf{3}, \mathbf{1})_0$ is required by the $P_{LR}$ symmetry.
This implies that additionally to the bidoublet $L$ and the $SU(2)_R$ triplet $E$ there will also be an $SU(2)_L$ triplet $E'$ appearing in the spectrum of composite resonances.
This choice of representations is in fact unique unless one allows for $SU(2)_R$ representations with dimension higher than 3 (which would imply the presence of states with exotic electric charges greater than $\pm2$).

The second generation lepton sector Lagrangian then reads
\begin{align}
\mathcal L_f =& \,\,
\bar l_L (i\slashed{\mathcal D}) l_L + \bar \mu_R (i \slashed{\mathcal D}) \mu_R \\
&+
\bar L (i\slashed{\mathcal D}-m_L) L
+
\bar E (i\slashed{\mathcal D}-m_E) E
+
\bar E' (i\slashed{\mathcal D}-m_E) E', \nonumber
\end{align}
where the covariant derivative $\mathcal D_\mu$ contains the couplings
to the composite vector resonances associated with the $SU(2)_L\times SU(2)_R\times
U(1)_X$ global symmetry\footnote{Contrary to \cite{Panico:2011pw}, we will
include resonances associated with $U(1)_X$ and $SU(3)_c$ in the following.} for the composite leptons and the coupling to the elementary gauge bosons for the elementary fermions.
The composite-elementary mixings can be written as~\footnote{In models where the Higgs boson is implemented as a pseudo Nambu-Goldstone boson these mixing terms correspond to an expansion in the Higgs non-linearities. For example, in a dimensionally deconstructed model like \cite{Panico:2011pw} with coset structure $SO(5)/SO(4)$ these are only the leading terms in $h/f$. In this case the composite fermions should be embedded into the $SO(5)$ adjoint representation $\mathbf{10}_0 = (\mathbf{2}, \mathbf{2})_0 \oplus (\mathbf{1}, \mathbf{3})_0 \oplus (\mathbf{3}, \mathbf{1})_0$ to achieve the custodial protection of the $Z$ vertex.}
\begin{align}
\mathcal L_\text{mix} &= 
    \lambda_L \, \text{tr}[ \bar\chi_L \, L_R  ]
  + \lambda_R \, \text{tr}[ \bar\chi_R \, E_L ] \nonumber \\
& \quad + Y_{L} \text{tr}[\bar L_L \mathcal{H} E_R] 
  + Y'_{L} \text{tr}[\mathcal{H} \bar L_L E'_R] \nonumber \\
& \quad + Y_{R} \text{tr}[\bar L_R \mathcal{H} E_L] 
  + Y'_{R} \text{tr}[\mathcal{H} \bar L_R E'_L] \nonumber \\
& \quad + \text{h.c.} \,
\end{align}
where $\chi_L$ and $\chi_R$ denote the embeddings of the SM leptons into $(\mathbf{2}, \mathbf{2})_0$ and $(\mathbf{1}, \mathbf{3})_0$, respectively, and $\mathcal{H}$ is the Higgs doublet transforming as a $(\mathbf{2}, \mathbf{2})_0$.

In the mass basis, we obtain a muon with mass
\begin{equation}
m_\mu = \frac{Y_{L}}{2 \sqrt{2}} \, \left< h \right>  \,
s_{L\mu}s_{R\mu} \,,
\label{eq:mmu}
\end{equation}
where $\left< h \right>$ is the VEV of the Higgs field and $s_{L,R}\equiv\sin\theta_{L,R}$ are the degrees of compositeness of left-
and right-handed muons, determined by ${\tan\theta_L=\lambda_L/m_L}$ and 
$\tan\theta_R=\lambda_R/m_E$.

At this point, the muon neutrino is still massless and we have not introduced
any mixing between the different lepton families to avoid constraints from
charged lepton flavour violating processes. We do not attempt to construct a
full model accounting for neutrino masses and mixing but instead focus on the constraints
on muon compositeness that are present even without lepton mixing.

\section{Numerical analysis}

\subsection{Quark flavour physics}

To generate a visible NP effect in the $b\to s\mu^+\mu^-$ transition,
there must be sufficiently large flavour violating interactions involving
left-handed quarks.
But apart from this requirement, other details of
the (composite) quark sector, such as the representations of composite quarks
or the presence of flavour symmetries or flavour anarchy, are not important for
our conclusions. This is because the same flavour-changing 
coupling that enters the $b\to s\,\ell^+\ell^-$ transition also enters
$B_s$-$\bar B_s$ mixing and is thus constrained from above.

\begin{figure}
\centering
\raisebox{0.75cm}{(a)}~~\includegraphics[width=0.7\columnwidth]{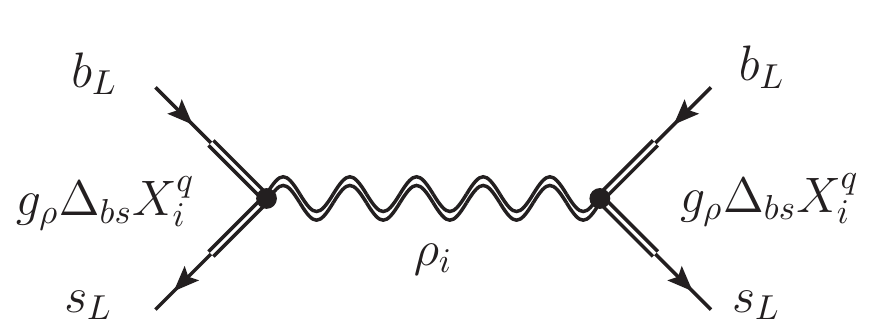}
\raisebox{0.75cm}{(b)}~~\includegraphics[width=0.7\columnwidth]{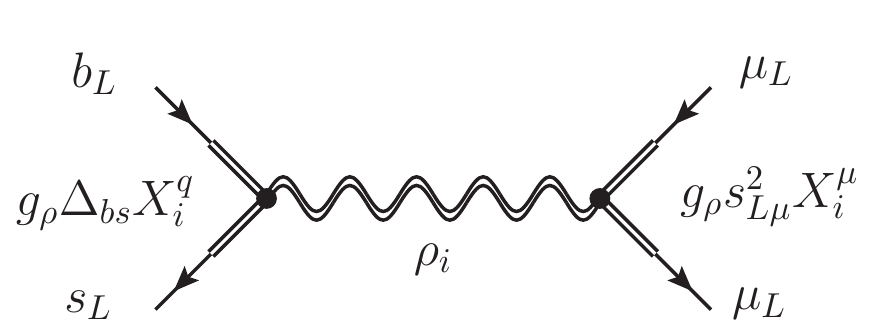}
\caption{Tree-level contribution to (a) $B_s$ mixing and (b)
$b\to s\mu^+\mu^-$ transitions. Double lines indicate composite fields,
$g_\rho$ is the coupling between composite fermion and vector resonances,
$s_{L\mu}$ the left-handed muons' degree of compositeness, $X_i^f$ is the charge
of the composite fermion mixing with $f$ under the global symmetry associated with
vector resonance $\rho_i$, and $\Delta_{bs}$ is a parameter depending on the
flavour structure and the degrees of compositeness of $b$ and $s$ quark.}
\label{fig:diagrams}
\end{figure}

The NP contribution to $B_s$-$\bar B_s$ mixing is encoded in the dimension-6
$\Delta B=2$ operator $O_V^{dLL}=(\bar s_{L}\gamma^\mu b_L)^2$
that arises from tree-level vector resonance exchange (see
fig.~\ref{fig:diagrams}a).
Its Wilson coefficient can be written as
\begin{equation}
C_V^{dLL} = \frac{g_\rho^2}{m_\rho^2} \,\Delta_{bs}^2 \,c_V^{dLL} \,,
\end{equation}
where $c_V^{dLL}$ is an $O(1)$ numerical factor that arises from the sum
over the quantum numbers of the composite quark partners under the global
symmetries associated with the exchanged vector resonances (indicated by
$X_i^q$ in fig.~\ref{fig:diagrams}a).
For both of the two choices of composite quark representations that 
feature a custodial protection of the $Z\bar b_L b_L$ coupling, one finds
$c_V^{dLL}=-23/36$ \cite{Barbieri:2012tu}.
The flavour violating parameter $\Delta_{bs}$ depends on the quark degrees of
compositeness, but a typical size (both in flavour anarchic models and in
models with a $U(2)^3$ flavour symmetry)
is $\mathcal{O}(1)\times V_{ts}^2$.

The Wilson coefficient of the $\Delta B=1$
operator $O_{dl}=(\bar s_{L}\gamma^\nu b_L)(\mu_L\gamma_\nu\mu_L)$,
that arises in an analogous way (see fig.~\ref{fig:diagrams}b), reads instead
\begin{equation}
C_{dl} = \frac{g_\rho^2}{m_\rho^2} \Delta_{bs} s_{L\mu}^2 \,c_{dl} \,,
\end{equation}
where $\Delta_{bs}$ is the same coupling as above and $c_{dl}=-1/2$ for our 
choice of representations.

We can then write a numerical formula for the deviation of $R_K$ from 1, as a 
function of the left-handed muons' degree of compositeness and the allowed 
deviation of the mass difference in $B_s$ mixing from the SM,
\begin{equation}
R_K - 1 \approx 
\pm 0.10
\left[\frac{1\,\text{TeV}}{f}\right]
\left[\frac{s_{L\mu}}{0.3}\right]^2
\left[\frac{|\Delta M_s-\Delta M_s^\text{SM}|}{0.1\,\Delta 
M_s^\text{SM}}\right]^{1/2}
\,,
\label{eq:RKth}
\end{equation}
where the negative sign holds for positive $\Delta_{bs}$
and we have used $m_\rho/g_\rho = f/2$ (see footnote~\ref{fn:f}).

Other constraints in the flavour sector, such as $K^0$-$\bar K^0$ mixing,
that typically represents a strong bound in models with flavour anarchy,
are more model-dependent. In models with Minimal Flavour Violation,
for instance, $b\leftrightarrow s$ transitions are the most constraining and
$K$ physics is not relevant in this respect.

\subsection{Electroweak precision constraints}

\begin{figure}[tbp]
\centering
\includegraphics[width=0.9\columnwidth]{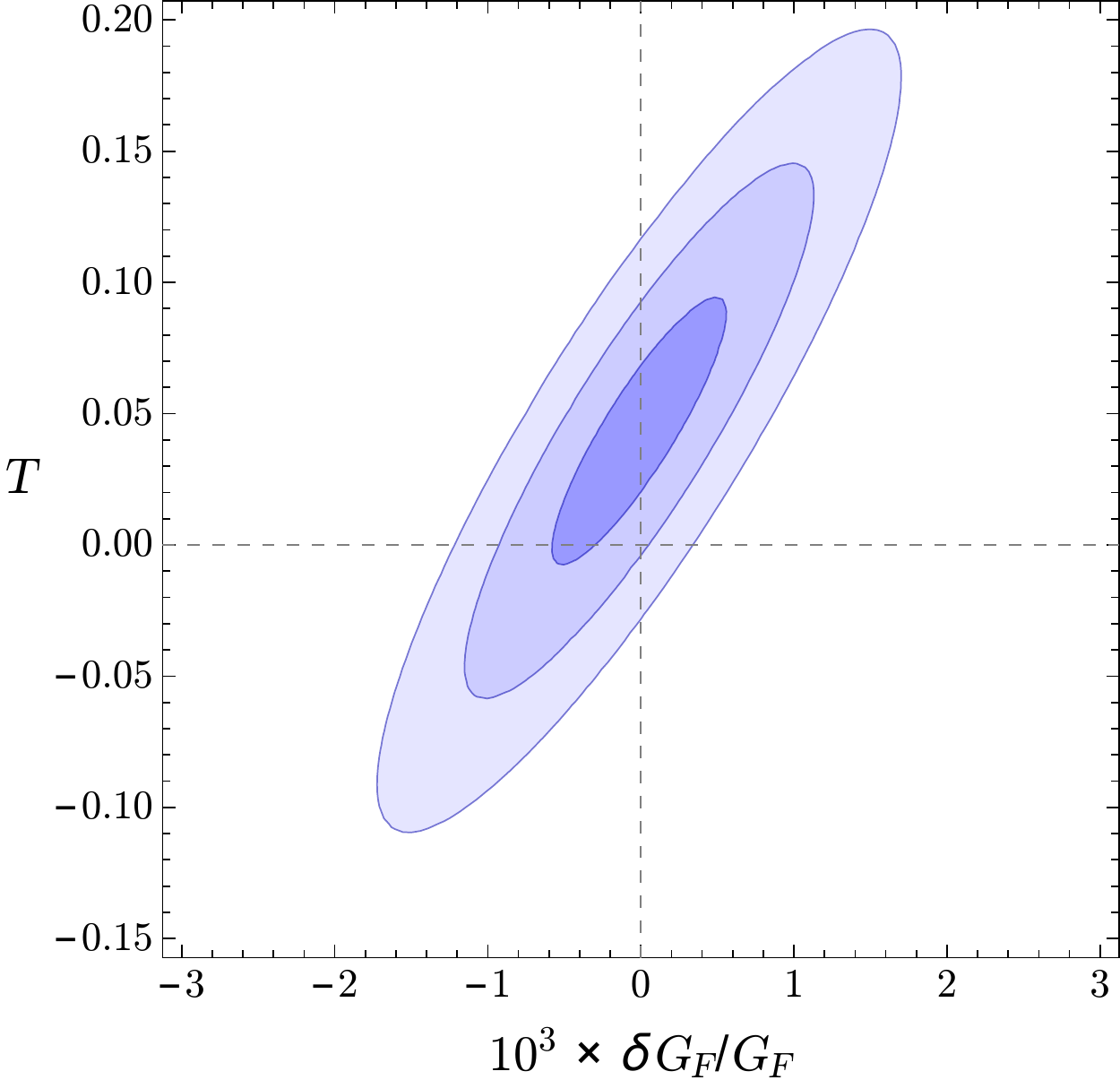}
\caption{Constraint at 1, 2, and $3\sigma$ on the modification of the Fermi
constant in muon decay with respect to the SM versus a NP contribution to
the electroweak $T$ parameter.}
\label{fig:ewp}
\end{figure}

\begin{figure}
\centering
\raisebox{0.75cm}{(a)}~~\includegraphics[width=0.7\columnwidth]{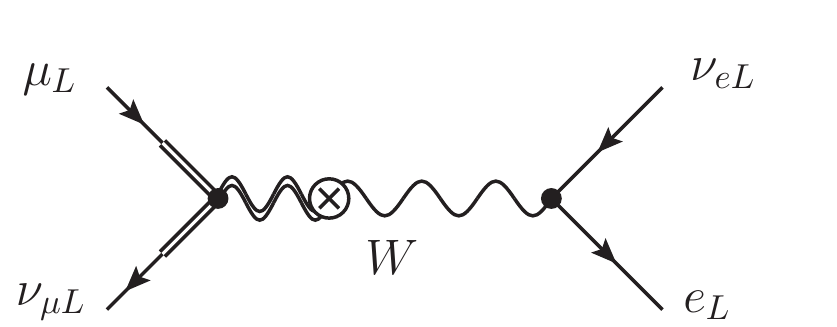}
\caption{Tree-level correction to the Fermi constant due to a shift
in the tree-level coupling $W\mu_L\nu_{\mu L}$ coupling. The circled cross
symbolized a double Higgs VEV insertion.}
\label{fig:diagramGF}
\end{figure}

Due to the discrete $P_{LR}$ symmetry of the fermion representations, 
the tree-level coupling of left-handed muons to the $Z$ boson is
custodially protected and thus SM-like by construction.
An additional loop-correction to this coupling might be relevant in a complete analysis of a specific model \cite{Grojean:2013qca}. However, this is beyond the scope of the present study whose intention is mainly a proof of concept. We will thus neglect the loop-contributions and focus solely on the tree-level effects.

In contrast to the neutral current coupling, the custodial protection
is not active for the charged current coupling $W\mu_L\nu_{\mu L}$.
A shift in this coupling would affect the muon lifetime and the extraction
of the Fermi constant.
To determine the allowed room for new physics in this coupling, a global
fit to electroweak precision observables has to be performed.
Importantly, the constraint on this coupling is strongly correlated
with the constraint on the electroweak $T$ parameter, which receives
loop contributions in composite Higgs models that depend on the details
of the quark sector.
Following \cite{Wells:2014pga}, we find the constraint shown in
fig.~\ref{fig:ewp}.
To leading order in $s_{L\mu}$ and $\xi$, the correction to the Fermi constant,
as illustrated by the diagram in fig.~\ref{fig:diagramGF},
reads
\begin{equation}
\frac{\delta G_F}{G_F} = \frac{\delta g^L_{W\mu\nu}}{g^L_{W\mu\nu}}=
-\frac{1}{4}\xi s_{L\mu}^2\left(1+\frac{m_L^2}{m_E^2}\right) \,.
\label{eq:Wmun}
\end{equation}
The first term in the bracket originates from the mixing of the
$W$ boson with the composite vector resonances, the second one from the mixing
of the left-handed leptons with the composite fermion triplets.
In the most favourable case of $m_L\ll m_E$ and a negative NP contribution
to the $T$ parameter, the constraint in fig.~\ref{fig:ewp} implies
$s_{L\mu}\lesssim 0.08\,\xi^{-1/2}$.

Likewise, the neutral current coupling to neutrinos, $Z\nu_{\mu L}\nu_{\mu L}$,
is not custodially protected either\footnote{We thank Gilad Perez for bringing this
point to our attention.}. To leading order in $\xi$,
the relative correction is equal to the relative correction to the
charged current up to a factor of 2,
\begin{equation}
\frac{\delta g^L_{Z\nu\nu}}{g^L_{Z\nu\nu}}=2\frac{\delta g^L_{W\mu\nu}}{g^L_{W\mu\nu}}
\,,
\label{eq:Znn}
\end{equation}
which is true for any model with a custodial protection of the 
$Z\mu_L\mu_L$ coupling (cf.\ \cite{Agashe:2006at}).
Experimentally, the $Z$ coupling to neutrinos is constrained by the invisible $Z$ width
measured at LEP, that can be expressed as the effective number of light neutrino species
$N_\nu$ via $\Gamma_\text{inv}=N_\nu\Gamma_{\nu\bar\nu}$. \eqref{eq:Znn} leads to a
modification of $N_\nu$,
\begin{equation}
N_\nu = 3 + 2 \frac{\delta g^L_{Z\nu\nu}}{g^L_{Z\nu\nu}} \,.
\end{equation}
Interestingly, the measurement of this quantity at LEP shows a $2\sigma$ deficit
\cite{ALEPH:2005ab},
\begin{equation}
N_\nu = 2.9840 \pm 0.0082 \,.
\label{eq:Nnuexp}
\end{equation}
We observe that in our setup, the correction always has negative sign and that the
maximum allowed value of \eqref{eq:Wmun} at $2\sigma$ according to
fig.~\ref{fig:ewp}, combined with \eqref{eq:Znn}, leads to a correction to $N_\nu$ that
basically coincides with \eqref{eq:Nnuexp}. We conclude that the tree-level correction
to the $Z\nu_{\mu L}\nu_{\mu L}$ coupling {\em improves} the agreement with the data.

Now, inserting the maximum value of $s_{L\mu}$ allowed by \eqref{eq:Wmun}
back into eq.~\eqref{eq:RKth}, one finds,
choosing the sign preferred by \eqref{eq:RK},
\begin{equation}
1-R_K\lesssim 0.12 \left[\frac{f}{1\,\text{TeV}}\right]
\left[\frac{|\Delta M_s-\Delta M_s^\text{SM}|}{0.1\,\Delta 
M_s^\text{SM}}\right]^{1/2}
\,.
\end{equation}
Consequently, the anomaly \eqref{eq:RK} can be explained at the $1\sigma$ level for
$f\gtrsim 1.3\,\text{TeV}$ and $s_{L\mu}\gtrsim 0.4$, at the same time explaining the
anomaly in the invisible $Z$ width at LEP.

\begin{figure}[tbp]
\centering
\includegraphics[width=0.95\columnwidth]{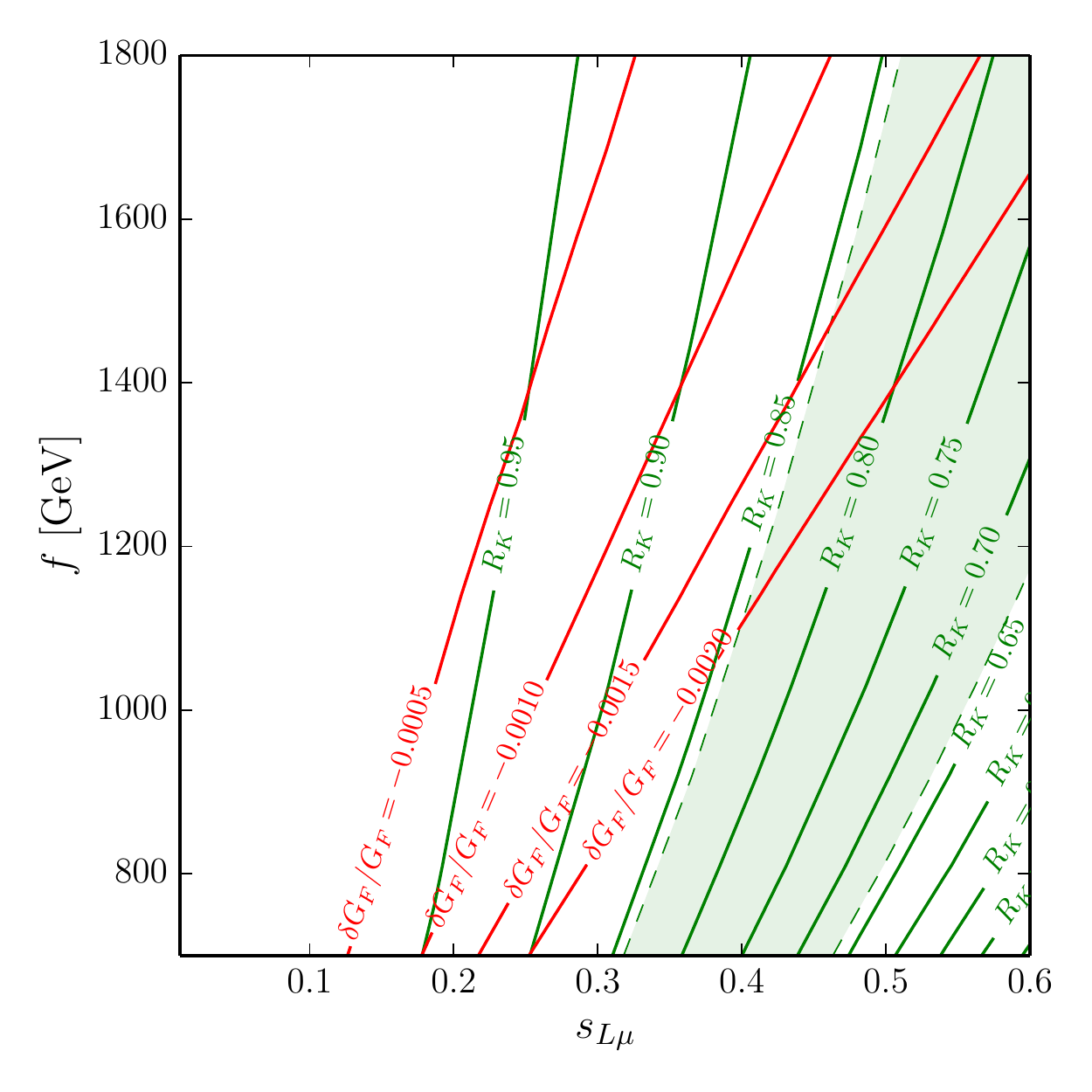}
\caption{Predictions for $R_K$ (green) and the relative shift in the Fermi constant (red) for a benchmark point with $m_L/m_E=0.3$. The flavour-changing coupling $\Delta_{bs}$ has been fixed to its maximum value allowing a 10\% shift in $\Delta M_s$. The green shaded region corresponds to the $1\sigma$ region allowed by \eqref{eq:RK}.
We do not show contours for $|{\delta G_F}/{G_F}|>0.002$, which is disfavoured (cf.~fig.~\ref{fig:ewp}).}
\label{fig:constraints}
\end{figure}

Fig.~\ref{fig:constraints} shows 
the values of ${\delta G_F}/{G_F}$ and of $R_K$ according to \eqref{eq:RKth}, assuming the flavour violating coupling $\Delta_{bs}$
to saturate a $10\%$ correction to $\Delta M_s$ and setting $m_L/m_E=0.3$.

\section{Conclusions}

We have demonstrated that a departure from lepton flavour universality in
$B^+\to K^+\ell^+\ell^-$ decays, as hinted by the recent LHCb measurement, 
could be explained
in minimal composite Higgs models if left-handed muons have a
sizable degree of compositeness. Assuming a generic composite Higgs with a global custodial symmetry $SO(4)$\footnote{But one should keep in mind that the results also remain valid for more realistic models e.g. with a pNG Higgs.}, the requirement to satisfy LEP
bounds on departures from lepton flavour universality in $Z\ell\ell$ couplings
uniquely fixes the representations of the composite lepton partners.
The strongest constraint on the model then comes from modified $W$ couplings.
Depending on the size of the loop corrections to the electroweak $T$ parameter,
a departure at the level of 10-20\% from $R_K=1$ is possible for $f\sim 1\,\text{TeV}$.
Interstingly, the correction also affects the $Z$ coupling to neutrinos and can
explain the deficit in the invisible $Z$ width observed at LEP in the same parameter
region that explains the $R_K$ anomaly.

If this model is realized in nature, there are several ways to test it beyond $R_K$.
\begin{itemize}
\item It predicts $\delta C_9^\mu \approx -\delta C_{10}^\mu$, which can be tested by
global fits to measurements of $b\to s$ transitions, including in particular
angular observables in ${B\to K^*\mu^+\mu^-}$.
This relation also implies a suppression of $B_s\to\mu^+\mu^-$ at the same
level of the suppression in $B^+\to K^+\mu^+\mu^-$ (cf.~\cite{Glashow:2014iga});
\item Deviations from LFU are also expected in other branching ratios and in the
forward-backward asymmetry in $B\to K^*\mu^+\mu^-$ at low $q^2$
(cf.~\cite{Altmannshofer:2014rta});
\item It implies an enhancement of both $B \to K \bar \nu \nu$ and $B\to K^* \bar \nu \nu$ correlated with, but roughly a factor of 5 smaller than, the suppression of $R_K$. Larger effects in these decays could be generated if taus are significantly composite as well (cf. \cite{Buras:2014fpa}); 
\item In principle, vector resonances could be too heavy to be in the reach
of the LHC. But if they are light enough, neutral electroweak resonances are
expected to have a sizable branching ratio into muons and could show up as
peaks in the dimuon invariant mass distribution;
\item The model predicts a positive contribution to the $B_s$ meson mass
difference $\Delta M_s$, which could be seen when the precision on the
relevant lattice parameters and the tree-level determination of the CKM
matrix improve in the future.
\end{itemize}

Finally, we note that our model is incomplete as it does not address neutrino masses
or give a rationale for the absence of charged lepton flavour violation. If
the anomaly~\eqref{eq:RK} holds up against further experimental scrutiny, it
will be interesting to investigate whether our model can be combined with a realistic
mechanism for lepton flavour.
Also, loop effects not considered in this letter may lead to additional constraints that should be included in a more complete analysis.

\subsection*{Acknowledgements}
This work was supported by the DFG cluster of excellence ``Origin and Structure of the Universe''.

\bibliography{compmuons}

\end{document}